\documentclass{article}
\usepackage{authblk}
\usepackage{geometry}
\usepackage{graphicx} 
\usepackage{amsmath}

\usepackage{amssymb}
\usepackage{multirow}
\usepackage{bm}
\usepackage{upgreek}
\usepackage[edges]{forest}
\usepackage{parskip}
\usepackage{orcidlink}
\usepackage{xcolor}
\usepackage[intoc]{nomencl}
\usepackage[toc,page]{appendix}
\usepackage{chngcntr}

\makenomenclature
\hypersetup{
    colorlinks,
    linkcolor={red!50!black},
    citecolor={blue!50!black},
    urlcolor={blue!80!black}
}

\sffamily

\title{The interplay between thermomigration and stress-driven hydrogen transport in metals\\\rule[0.2cm]{15cm}{0.4pt}}
\author[]{Daniel J. Long$^*$\orcidlink{0009-0004-2068-7871}}
\author[]{Edmund Tarleton\orcidlink{0000-0001-6725-9373}}
\author[]{Alan C.F. Cocks\orcidlink{0000-0002-6245-3406}}
\author[]{Felix Hofmann$^{\dag}$\orcidlink{0000-0001-6111-339X}}
\affil[]{Department of Engineering Science, University of Oxford, Parks Road, OX1 3PJ Oxford, UK}
\begin{document}
\date{}
\maketitle
\centerline{Correspondence: $^*$\href{mailto:daniel.long@eng.ox.ac.uk}{daniel.long@eng.ox.ac.uk}, $^{\dag}$\href{mailto:felix.hofmann@eng.ox.ac.uk}{felix.hofmann@eng.ox.ac.uk}}



\nomenclature{$\boldsymbol{\bar{J}}_{\mathrm{H}}$}{Hydrogen flux vector}
\nomenclature{$\boldsymbol{J}_{\mathrm{q}}$}{Heat flux vector}

\nomenclature{$D_{\mathrm{L}}$}{Lattice hydrogen diffusivity}
\nomenclature{$\bar{C}_{\mathrm{L}}$}{Lattice hydrogen concentration}
\nomenclature{$\theta_{\mathrm{L}}$}{Lattice hydrogen occupancy}

\nomenclature{$\mu$}{Chemical potential of hydrogen}
\nomenclature{$\mu_{\mathrm{eff}}$}{Effective chemical potential of hydrogen}
\nomenclature{$\mu_0$}{Reference chemical potential}

\nomenclature{$\sigma_{\mathrm{H}}$}{Hydrostatic stress}
\nomenclature{$V_{\mathrm{L}}$}{Partial molar volume of hydrogen in the lattice}

\nomenclature{$T$}{Absolute temperature}

\nomenclature{$Q^*$}{Heat of transport}
\nomenclature{$Q^*_{\mathrm{int}}$}{Intrinsic contribution to the heat of transport}
\nomenclature{$Q^*_{\mathrm{es}}$}{Electrostatic contribution to the heat of transport}
\nomenclature{$Q^*_{\mathrm{ele}}$}{Electron-wind contribution to the heat of transport}
\nomenclature{$Q^*_{\sigma_{\mathrm{H}}=0}$}{Heat of transport in the absence of hydrostatic stress}
\nomenclature{$q_i$}{Coefficients in the polynomial representation of $Q^*_{\sigma_{\mathrm{H}}=0}$}
\nomenclature{$Z_{\mathrm{es}}$}{Charge transfer coefficient between the lattice and hydrogen}

\nomenclature{$\varepsilon_{ii}^{\mathrm{el}}$}{Principal elastic strains}
\nomenclature{$\varepsilon_{ii}^{\mathrm{th}}$}{Thermal strains}
\nomenclature{$\sigma_{ii}$}{Principal stresses}

\nomenclature{$h_{\mathrm{vib}}$}{Specific vibrational enthalpy of hydrogen}
\nomenclature{$\omega_i$}{Hydrogen vibrational frequency}
\nomenclature{$\bar{\omega}_i$}{Hydrogen vibrational wave number}
\nomenclature{$g_i$}{Degeneracy of hydrogen vibrational mode}
\nomenclature{$c$}{Speed of light in vacuum}
\nomenclature{$M$}{Number of vibrational modes}

\nomenclature{$R$}{Universal gas constant}
\nomenclature{$k_{\mathrm{B}}$}{Boltzmann constant}
\nomenclature{$N_{\mathrm{A}}$}{Avogadro constant}
\nomenclature{$\hbar$}{Reduced Planck constant}

\nomenclature{$L_{ij}$}{Onsager transport coefficients}
\nomenclature{$\boldsymbol{X}_{\mathrm{H}}$}{Thermodynamic driving force for hydrogen transport}
\nomenclature{$\boldsymbol{X}_{\mathrm{q}}$}{Thermodynamic driving force for heat transport}

\nomenclature{$k$}{Thermal conductivity}
\nomenclature{$\alpha_{\mathrm{S}}$}{Seebeck coefficient}
\nomenclature{$\alpha_{\mathrm{T}}$}{Thomson coefficient}
\nomenclature{$\rho$}{Electrical resistivity}
\nomenclature{$\rho_{\mathrm{th}}$}{Thermal contribution to electrical resistivity}
\nomenclature{$N_{\mathrm{H}}$}{Hydrogen number density}
\nomenclature{$N_{\mathrm{e}}$}{Free electron number density}
\nomenclature{$\gamma,\delta$}{Electron and hole band contribution parameters}

\nomenclature{$E$}{Young’s modulus}
\nomenclature{$\nu$}{Poisson’s ratio}
\nomenclature{$\alpha_{\mathrm{th}}$}{Coefficient of thermal expansion}

\nomenclature{$h$}{Convective heat transfer coefficient at a surface}
\nomenclature{$T_{\mathrm{env}}$}{Environmental temperature at a boundary}
\nomenclature{$T_{\mathrm{s}}$}{Surface temperature at a boundary}
\nomenclature{$D_0$}{Diffusion pre-exponential factor}
\nomenclature{$Q_{\mathrm{D}}$}{Activation energy for hydrogen diffusion}
\nomenclature{$c_\mathrm{p}$}{Specific heat capacity at constant pressure}
\nomenclature{$f$}{Thermomigration contribution to the effective chemical potential}


\printnomenclature

\section*{Abstract}

Thermomigration is the driving force for hydrogen transport due to a temperature gradient. It can compete with hydrogen transport induced by stress gradients. While stress-driven hydrogen migration is well established, thermomigration remains comparatively underexplored, largely due to limited mechanistic understanding and a scarcity of experimental data. In this work, we develop a thermodynamically consistent framework for hydrogen transport, incorporating a mechanistic model for thermomigration. This is implemented within a finite element framework using an effective chemical potential. Using case studies of iron and nickel heat exchangers and zirconium alloy nuclear fuel cladding, we quantify the competing and synergistic effects of thermomigration and stress-driven transport. We show that thermomigration often dominates hydrogen redistribution in heat-carrying components, even in the presence of significant thermal incompatibility stresses. However, stress-driven transport is shown to become decisive near sharp stress concentrators. A graphical method is introduced to rapidly identify the dominant transport mechanism without requiring fully coupled simulations. The results provide practical guidance for assessing hydrogen redistribution and embrittlement risk in heat-carrying structural components.

\section{Introduction}

A fundamental understanding of the driving forces for hydrogen (H) transport is essential for developing robust failure criteria for structural components exposed to H. This is because of H-linked degradation phenomena in metals, often collectively categorised under the umbrella term of hydrogen embrittlement (HE) \cite{DWIVEDI2018,SUN2023}. Among the most widely reported HE mechanisms are hydrogen-enhanced localised plasticity (HELP) and hydrogen-enhanced decohesion (HEDE). The underlying theory for HELP is that dislocations represent energetically favourable sites for H (relative to the pristine lattice) and hence act as traps \cite{WERT2005}. Trapped H can effectively shield the elastic strain fields of dislocations \cite{FERREIRA1998,SOFRONIS1995,YANG2025}, enhancing their mobility, thus leading to enhanced plastic strain localisation and damage accumulation \cite{VON2011}. The mechanistic basis for HEDE requires that interfaces such as grain boundaries and phase boundaries also act as traps \cite{HE2021}. Experimental measurements \cite{WASIM2021,BECHTLE2009,KAMEDA1983} and ab initio models \cite{MCENIRY2018,YAMAGUCHI2019,WILSON2022} have shown that H trapped at these interfaces can severely degrade their cohesive strengths, leading to premature decohesion and embrittlement. Both mechanisms are highly sensitive to the trapped H concentration \cite{OKUNO2023}, which depends on the transport of mobile H via the lattice \cite{KEHR2005}. The development of concentration-dependent failure criteria for HE also therefore requires a strong understanding of the interactions between surfaces and their H-rich environments (for reliable H uptake predictions). In this paper, we focus on the driving forces for internal H redistribution only. 

In engineering applications where components are subjected to complex thermomechanical loading histories, a number of different thermodynamic forces can drive the transport of H within the body. The primary driving forces for H transport in metals originate from gradients of (i) H concentration, (ii) stress, (iii) temperature (thermomigration), and (iv) electrical potential (electromigration) \cite{ADLER1978,WIPF2005}. The well-known Fick's first diffusion law characterises (i) \cite{PHILIBERT2006}, while models for (ii) are derived from equilibrium between elastic and H swelling strain energies \cite{SOFRONIS1995}. Stress gradient effects on H localisation and embrittlement have been widely investigated in the literature, with studies exploring the influence of heterogeneity from microstructural \cite{LONG2025} to continuum \cite{ELMUKASHFI2020} (geometric) length scales. Driving forces (i) and (ii) are hence well understood. By contrast, very few studies since the Space Race have seriously explored (iii). Until recently, thermomigration modelling efforts have been confined to the use of scarcely available empirical data \cite{GONZALEZ1965} due to limited mechanistic understanding. We recently addressed this knowledge gap \cite{LONG2025THERMO} using a nonequilibrium thermodynamic framework, the details of which are summarised here in Sections \ref{Q*} and \ref{framework}. The quantity used to characterise the magnitude and direction of H flux (or particle flux, more generally) in a temperature gradient is the heat of transport, $Q^*$. When $Q^*$ is positive, the thermomigration direction matches that of the heat flux, i.e., down the temperature gradient, and vice versa \cite{CHEN2012}. Zhang et al. \cite{ZHANG2022} studied the role of temperature, stress, and the gradients thereof on the localisation and trapping of H in nickel alloys for H gas turbine applications using finite element analyses. Thermomigration was shown to dominate over stress-driven transport of diffusible H, leading to localisation by up to two orders of magnitude, while the latter was responsible for localisation of less than 10\%. Elevated temperatures were also shown to accelerate the trapping rate of H at dislocations due to increased levels of plasticity. A key shortcoming of this study however was that it relied on a single point experimental measurement of $Q^*$ (which was obtained at a fixed temperature) for analyses across a broad temperature range. We know from the limited available experimental data \cite{GONZALEZ1965} and from our model \cite{LONG2025THERMO} that $Q^*$ is highly temperature dependent. Moreover, while the study usefully showed that thermomigration was the dominant driving force for specific thermomechanical loading conditions, the results are difficult to generalise, particularly for understanding systems from outside the model bounds. Electromigration is not discussed here because it is generally not relevant for applications in which structural integrity is a concern. 

As industries transition towards sustainable energy, the role of H is set to become increasingly important. In aviation, H-powered jet engine technology represents a promising low-carbon alternative to conventional propulsion systems, provided that a low-cost, fully green H supply chain can be established \cite{ADLER2023AIRCRAFT}. A major challenge for the industry however is the meagre volumetric energy density offered by gaseous H, compared with current fuels \cite{VERSTRAETE2015}. To mitigate this limitation, H aviation fuel is expected to be stored in cryogenic vessels in liquid form, thereby partially increasing its volumetric energy density. For engineers, this introduces a major new challenge: between storage and combustion, pre-heating will be required to vaporise liquid H to ensure steady, isothermal, single-phase flow into the combustor \cite{ROMPOKOS2025}. A heat exchanger device may be used to facilitate H pre-heating. Hence, this represents a new application in which the redistribution and localisation of H is dependent on both stress-migration (as incompatibility stresses arise in heat exchangers \cite{MIAO2017}) and thermomigration driving forces. Such examples are often found in heat-carrying components. In the nuclear energy industry, for example, zirconium alloy reactor fuel cladding materials are exposed to temperature gradients, thermal incompatibility stresses, and H (from oxidation in water) \cite{COUET2014}. The aim of this paper is to explore the conditions that promote the dominance of either H transport driving force. Using heat exchanger and reactor fuel cladding case studies, we demonstrate a general and computationally efficient graphical approach for the swift analysis of any system.

\section{Methodology}
\subsection{Hydrogen transport model} \label{JHmodel}

In isothermal studies of stress-driven H transport, only concentration and stress gradient contributions to the H flux, $\boldsymbol{\bar{J}}_{\mathrm{conc}}$ and $\boldsymbol{\bar{J}}_{\mathrm{stress}}$, respectively, are accounted for. From Elmukashfi et al. \cite{ELMUKASHFI2020}, both driving forces are considered using a model based on thermodynamic equilibrium, i.e. via the gradient of chemical potential of H. Hence, the H flux is given by $\boldsymbol{\bar{J}}_{\mathrm{H}}=-\frac{D_{\mathrm{L}}\bar{C}_{\mathrm{L}}}{RT}\nabla\mu$, where $D_{\mathrm{L}}$ is H diffusivity, $\bar{C}_{\mathrm{L}}$ the lattice (diffusible) molar H concentration, $R$ the gas constant, $T$ is temperature, $\mu=\mu_0-V_{\mathrm{L}}\sigma_{\mathrm{H}}+RT\ln{(\theta_{\mathrm{L}})}$, $\mu_0$ is the 'standard' chemical potential, defined at some reference temperature and pressure, $V_{\mathrm{L}}$ is the partial molar volume of H in the lattice, $\sigma_{\mathrm{H}}$ is the hydrostatic component of stress, and $\theta_{\mathrm{L}}$ is the H lattice occupancy. The lattice H concentration describes the absolute amount of H present in the lattice, while the lattice occupancy represents the fraction of available lattice sites that are occupied by H, indicating the level of saturation. This gives $\boldsymbol{\bar{J}}_{\mathrm{H}}=-\frac{D_{\mathrm{L}}\bar{C}_{\mathrm{L}}}{RT}\left(-V_{\mathrm{L}}\nabla\sigma_{\mathrm{H}}+RT\nabla\ln{(\theta_{\mathrm{L}})}\right)$. When there is a temperature gradient, the thermomigration contribution, $\boldsymbol{\bar{J}}_{\mathrm{thermo}}$, is accounted \cite{ZHANG2022} for to give

\begin{equation}\label{JH1}
    \boldsymbol{\bar{J}}_{\mathrm{H}}=-\frac{D_{\mathrm{L}}\bar{C}_{\mathrm{L}}}{RT}\left(-V_{\mathrm{L}}\nabla\sigma_{\mathrm{H}}+RT\nabla\ln{(\theta_{\mathrm{L}})}+\frac{Q^*}{T}\nabla T\right),
\end{equation}

where $\boldsymbol{\bar{J}}_{\mathrm{H}}=\boldsymbol{\bar{J}}_{\mathrm{stress}}+\boldsymbol{\bar{J}}_{\mathrm{conc}}+\boldsymbol{\bar{J}}_{\mathrm{thermo}}$ and $Q^*$ is the heat of transport. To reliably predict thermomigration behaviour across a broad temperature range, a mechanistic heat of transport model is used. This is described in the next sub-section.  

\subsection{Heat of transport model} \label{Q*}

\begin{figure}[ht]
\includegraphics[width=1\linewidth]{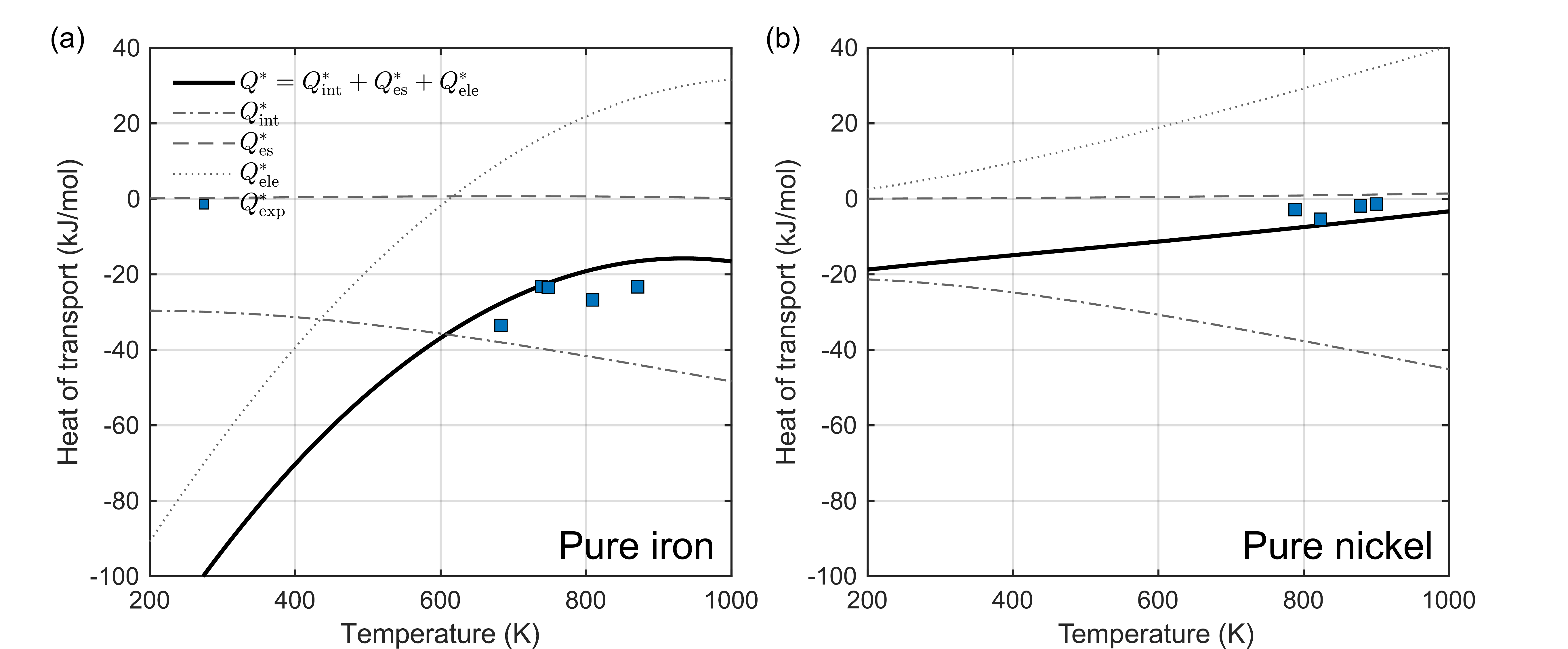}
\caption{\label{Qstar_comparison} Comparison of mechanistic model for the heat of transport with experimental data, $Q^*_{\mathrm{exp}}$, from Gonzalez and Oriani \cite{GONZALEZ1965} for (a) iron and (b) nickel. Dashed lines represent individual contributions (intrinsic, electrostatic, and electron-wind) to the total heat of transport.}
\end{figure}

We previously showed that in metals with cubic symmetry, the heat of transport is given by the sum of three distinct contributions \cite{LONG2025THERMO},

\begin{equation}\label{Q*1}
    Q^*=Q^*_{\mathrm{int}}+Q^*_{\mathrm{es}}+Q^*_{\mathrm{ele}},
\end{equation}

\noindent where $Q^*_{\mathrm{int}}$ is the intrinsic contribution, $Q^*_{\mathrm{es}}$ the electrostatic contribution, and $Q^*_{\mathrm{ele}}$ derives from an electron-wind effect. The intrinsic contribution stems from differences in H hopping rates between atomic lattice sites along the temperature gradient. It is given by 

\begin{equation}\label{Q*int}
    Q^*_{\mathrm{int}}=-h_{\mathrm{vib}}+V_{\mathrm{L}}\sigma_{\mathrm{H}},
\end{equation}

\noindent where $h_{\mathrm{vib}}$ is the specific vibrational enthalpy of H, $V_{\mathrm{L}}$ the partial molar volume of H in the lattice, and $\sigma_{\mathrm{H}}$ is the internal hydrostatic stress. For most materials, $Q^*_{\mathrm{int}}$ is likely to be dominated by $h_{\mathrm{vib}}$ \cite{DALARSSON2011}, which is given by

\begin{equation}\label{hvib}
    h_{\mathrm{vib}}=N_{\mathrm{A}}\sum_{i=1}^{M}\left[g_i\hbar\omega_i\left(\frac{1}{2}+\left(\exp{\left(\frac{\hbar\omega_i}{k_{\mathrm{B}}T}\right)}-1\right)^{-1}\right)\right],
\end{equation}

\noindent where $N_{\mathrm{A}}$ is Avogadro's constant, $g_i$ the degeneracy of vibrational mode $i$, $M$ the total number of vibrational modes, $\hbar$ is reduced Planck's constant, $\omega_i$ the H atom vibrational frequency, and $k_{\mathrm{B}}$ is Boltzmann's constant. In experiments, $\omega_i$, is often given in terms of the wave number \cite{KOLESOV2017}, $\bar{\omega}_i=\omega_i/c$, where $c$ is the speed of light in a vacuum. The electrostatic contribution to $Q^*$ arises due to the electrostatic interactions between H and thermoelectric fields, and is given by

\begin{equation}\label{Q*es}
    Q^*_{\mathrm{es}}=-Z_{\mathrm{es}}FT\alpha_{\mathrm{T}},
\end{equation}

\noindent where $Z_{\mathrm{es}}$ is the charge transfer coefficient between the lattice and H, $F$ is Faraday's constant, and $\alpha_{\mathrm{T}}$ is the Thomson coefficient. Lastly, a mechanistic model for the electron-wind contribution to $Q^*$ (in metals with electron and hold bands), based on momentum transfer during electron scattering due to H, was originally presented by Huntington \cite{HUNTINGTON1968}. Our revised version of this model \cite{LONG2025THERMO} is given by

\begin{equation}\label{Q*ele}
    Q^*_{\mathrm{ele}}=-FT\alpha_{\mathrm{S}}\frac{1+\gamma}{1-2\gamma}(1+3\delta)\frac{\partial\rho}{\partial N_{\mathrm{H}}}\cdot\frac{N_{\mathrm{e}}}{\rho_{\mathrm{th}}},
\end{equation}

\noindent where $\alpha_{\mathrm{S}}$ is the Seebeck coefficient, $\rho$ the electrical resistivity, $\rho_{\mathrm{th}}$ the contribution to resistivity arising from thermal scattering processes, $N_{\mathrm{H}}$ and $N_{\mathrm{e}}$ the number density of H (concentration with units atoms$\cdot$m$^{-3}$) and free electrons, respectively, and $\gamma$ and $\delta$ are dimensionless quantities which represent the ratio of the hole to electron band contribution to the electron drag force and electric current, respectively. The resultant temperature-dependent model for $Q^*$ is plotted and compared with experiments \cite{GONZALEZ1965} in Figure \ref{Qstar_comparison} for pure iron and pure nickel. All property data for this model were previously reported in \cite{LONG2025THERMO} and are provided here in Table \ref{Mechanistic_props}.

\subsection{A nonequilibrium thermodynamic framework for hydrogen transport} \label{framework}

For non-isothermal problems, thermodynamic equilibrium cannot be enforced. Hence, the chemical potential alone cannot adequately account for driving forces (i) - (iii). Moreover, in systems with concurrent transport processes (here, we have H and thermal transport), it is necessary to consider potential reciprocal interactions between them. From Onsager \cite{ONSAGERI1931,ONSAGERII1931}, we write

\begin{equation}\label{Onsager}
    \begin{pmatrix}
    \boldsymbol{\bar{J}}_{\mathrm{H}}\\
    {\boldsymbol{J}}_{\mathrm{q}}
    \end{pmatrix}=
    \begin{pmatrix}
    L_{\mathrm{HH}} & L_{\mathrm{Hq}}\\
    L_{\mathrm{qH}} & L_{\mathrm{qq}}
    \end{pmatrix}
    \begin{pmatrix}
    \boldsymbol{X}_{\mathrm{H}}\\
    \boldsymbol{X}_{\mathrm{q}}
    \end{pmatrix}\\,
\end{equation}

where $\boldsymbol{\bar{J}}_{\mathrm{H}}$ and ${\boldsymbol{J}}_{\mathrm{q}}$ are H and heat fluxes, respectively, $\boldsymbol{X}_{\mathrm{H}}=-\nabla(\frac{\mu}{T})$ and $\boldsymbol{X}_{\mathrm{q}}=\nabla\left(\frac{1}{T}\right)$ are the thermodynamic driving forces for chemical and thermal transport, respectively, and $L_{ij}$ are the Onsager coefficients. In nonequilibrium thermodynamics, the $\nabla\left(\frac{\mu}{T}\right)$ driving force is based on maximising entropy production, rather than energy minimisation via $\nabla\mu$. Note that the off-diagonal coefficients, $L_{\mathrm{Hq}}$ and $L_{\mathrm{qH}}$ relate H and thermal transport to thermal and H transport driving forces, respectively. From \cite{LONG2025THERMO}, we show that $L_{\mathrm{HH}}=\frac{D_{\mathrm{L}}\bar{C}_{\mathrm{L}}}{R}$ and $L_{\mathrm{Hq}}=L_{\mathrm{HH}}Q^*_{\mathrm{ele}}$, which gives

\begin{equation}\label{JH2}
    \boldsymbol{\bar{J}}_{\mathrm{H}}=-\frac{D_{\mathrm{L}}\bar{C}_{\mathrm{L}}}{R}\left(\nabla\left(\frac{\mu}{T}\right)+\frac{Q^*_{\mathrm{ele}}}{T^2}\nabla T\right).
\end{equation}

Hence, we highlight an important discrepancy between $Q^*_{\mathrm{ele}}$ and other contributions to $Q^*$. The electron-wind contribution is a secondary effect that arises from electron scattering due to H. Onsager's reciprocal condition enforces that $L_{\mathrm{qH}}=L_{\mathrm{Hq}}$, such that the effect of H electron scattering on thermal transport is also accounted for. From Equation (\ref{Onsager}), we have

\begin{equation}\label{Jq1}
    \boldsymbol{{J}}_{\mathrm{q}}=-\frac{D_{\mathrm{L}}\bar{C}_{\mathrm{L}}Q^*_{\mathrm{ele}}}{R}\nabla\left(\frac{\mu}{T}\right)-k\nabla T,
\end{equation}

where $k$ is the H-unaffected thermal conductivity. While it is shown that acounting for the electron-wind is vital for accurately quantifying thermomigration, the following question is raised: how important is the reciprocal effect for thermal transport modelling? A rough quantitative estimate can be very informative. For an isothermal system, we have $\boldsymbol{\bar{J}}_{\mathrm{H}}=-\frac{D_{\mathrm{L}}\bar{C}_{\mathrm{L}}}{RT}\nabla\mu$, such that the heat flux required to maintain isothermal conditions is $\boldsymbol{{J}}_{\mathrm{q}}=L_{\mathrm{qH}}\boldsymbol{X}_{\mathrm{H}}=\boldsymbol{\bar{J}}_{\mathrm{H}}Q^*_{\mathrm{ele}}$. Using large, conservative estimates for H flux, ${\bar{J}}_{\mathrm{H}}=10^{-7}$ mol/m$^2$s \cite{ADDACH2009} (in iron in 1D) and $Q^*_{\mathrm{ele}}=10^5$ J/mol \cite{LONG2025THERMO} yields a heat flux, ${{J}}_{\mathrm{q}}=10^{-2}$ W/m$^2$. This is equivalent to a heat flux driven by a temperature gradient of around $10^{-4}$ K/m in iron, which is negligible in this context, i.e. $\boldsymbol{{J}}_{\mathrm{q}}\approx-k\nabla T$. The reciprocal effect of H on thermal transport is hence not further considered in this paper. 

\subsection{Finite element implementation} \label{FEdetails}
\subsubsection{Incorporating the heat of transport within an effective chemical potential}\label{UMATHT}

In the H transport finite element modelling framework presented by Elmukashfi et al \cite{ELMUKASHFI2020}, H flux was given by $\boldsymbol{\bar{J}}_{\mathrm{H}}=-\frac{D_{\mathrm{L}}\bar{C}_{\mathrm{L}}}{RT}\nabla\mu$. The contributions from concentration and stress gradients are hence accounted for within a single term, $\nabla\mu$. Adapting the framework provided by the user-defined thermal material behaviour subroutine (UMATHT) in \textit{Abaqus} (originally for thermal transport modelling), Elmukashfi et al. \cite{ELMUKASHFI2020} could efficiently model H transport, using $\mu$ as a surrogate for $T$, and $\boldsymbol{\bar{J}}_{\mathrm{H}}$ as a surrogate for $\boldsymbol{{J}}_{\mathrm{q}}$. This approach is particularly advantageous for computing gradients of stress, since this would otherwise require computing the second order derivatives of displacement using non-linear elements. With the UMATHT code, linear elements may be reliably used. In an equivalent nonequilibrium framework, we have Equation (\ref{JH2}) for $\boldsymbol{\bar{J}}_{\mathrm{H}}$, which depends upon gradients of both $\frac{\mu}{T}$ and $T$. In this way, Equation (\ref{JH2}) is incompatible with the UMATHT framework. Hence, here we present a revised H flux model based on the gradient of $\frac{\mu_{\mathrm{eff}}}{T}$ alone, where $\mu_{\mathrm{eff}}$ is an effective chemical potential. H flux is now

\begin{equation}\label{JH3}
    \boldsymbol{\bar{J}}_{\mathrm{H}}=-\frac{D_{\mathrm{L}}\bar{C}_{\mathrm{L}}}{R}\nabla\left(\frac{\mu_{\mathrm{eff}}}{T}\right).
\end{equation}

Here we define the effective chemical potential of H,

\begin{equation}\label{mueff}
    \mu_{\mathrm{eff}}=-V_{\mathrm{L}}\sigma_{\mathrm{H}}+RT\ln{(\theta_{\mathrm{L}})}+f,
\end{equation}

where $f$ is some function of $T$, and incorporates the absent contributions from $Q^*$, such that Equations (\ref{JH2}) and (\ref{JH3}) are equivalent to one another. Note that here we omit $\mu_0$, since it is normally defined at constant temperature and therefore contributes only a reference term; in the present nonequilibrium setting with spatially varying $T$, its effects are instead incorporated into the function $f(T)$, which accounts for the temperature-dependent contributions to $\mu_{\mathrm{eff}}$. Using the gradient product rule, this new driving force is expanded to give

\begin{equation}\label{gradmueff1}
    \nabla\left(\frac{\mu_{\mathrm{eff}}}{T}\right)=\frac{1}{T}\nabla\mu_{\mathrm{eff}}-\frac{\mu_{\mathrm{eff}}}{T^2}\nabla T.
\end{equation}

Combining Equations (\ref{mueff}) and (\ref{gradmueff1}), we have 

\begin{equation}\label{gradmueff2}
    \nabla\left(\frac{\mu_{\mathrm{eff}}}{T}\right)=\frac{1}{T}\left(-V_{\mathrm{L}}\nabla\sigma_{\mathrm{H}}+RT\nabla\ln{(\theta_{\mathrm{L}})}+\frac{V_{\mathrm{L}}\sigma_{\mathrm{H}}}{T}\nabla T+\frac{\partial f}{\partial T}\nabla T-\frac{f}{T}\nabla T\right).
\end{equation}

To establish a relationship between $f$ and $Q^*$, we combine Equations (\ref{JH3}) and (\ref{gradmueff2}) for $\boldsymbol{\bar{J}}_{\mathrm{H}}$ and let this equal Equation (\ref{JH1}). To satisfy this constraint, we have the following expression for $Q^*$ in terms of $f$,

\begin{equation}\label{Q*2}
    Q^*=V_{\mathrm{L}}\sigma_{\mathrm{H}}+T\frac{\partial f}{\partial T}-f.
\end{equation}

Note that the contribution to $Q^*_{\mathrm{int}}$ from hydrostatic stress emerges here in this derivation. Hence, $f$ depends only on the stress-independent heat of transport, $Q^*_{\sigma_{\mathrm{H}}=0}=Q^*-V_{\mathrm{L}}\sigma_{\mathrm{H}}$. We may then rearrange Equation (\ref{Q*2}) to write

\begin{equation}
    \frac{Q^*_{\sigma_{\mathrm{H}}=0}}{T^2}=\frac{1}T{}\frac{\partial f}{\partial T}-\frac{f}{T^2}=\frac{\partial}{\partial T}\left(\frac{f}{T}\right).
\end{equation}

We may now write the following general expression for $f$ in terms of $Q^*_{\sigma_{\mathrm{H}}=0}$,

\begin{equation}\label{f1}
    f=T\int\frac{Q^*_{\sigma_{\mathrm{H}}=0}}{T^2}dT.
\end{equation}

Whilst we are now in a position to solve this integral using Equation (\ref{Q*1}) as an input, for pragmatic reasons, we instead represent the temperature dependence of $Q^*$ using a second-order polynomial fit to Equation (\ref{Q*1}). Hence, we write $Q^*_{\sigma_{\mathrm{H}}=0}=q_1T^2+q_2T+q_3$, where $q_i$ are fitted polynomial coefficients. Equation (\ref{f1}) is now

\begin{equation}\label{f2}
    f=q_1T^2+q_2T\ln{(T)}-q_3+AT,
\end{equation}

and $A$ is an integration constant. Hence, within the UMATHT code, the new surrogate term for the temperature variable is given by

\begin{equation}\label{mueff_T}
    \frac{\mu_{\mathrm{eff}}}{T}=-\frac{V_{\mathrm{L}}\sigma_{\mathrm{H}}}{T}+R\ln{(\theta_{\mathrm{L}})}+q_1T+q_2\ln{(T)}-\frac{q_3}{T}+A.
\end{equation}

In the actual implementation, the constant, $A$, is omitted from Equation (\ref{mueff_T}), since $\nabla A=0$.

\subsubsection{Decoupled thermal transport and hydrogen transport analyses}

As discussed, the UMATHT subroutine in \textit{Abaqus} is used to model H transport problems via $\boldsymbol{\bar{J}}_{\mathrm{H}}=-\frac{D_{\mathrm{L}}\bar{C}_{\mathrm{L}}}{R}\nabla\left(\frac{\mu_{\mathrm{eff}}}{T}\right)$, with degree of freedom $\frac{\mu_{\mathrm{eff}}}{T}$, because it is directly analogous to Fourier's law for conduction heat transfer, $\boldsymbol{J}_{\mathrm{q}}=-k\nabla T$, with degree of freedom $T$. Generally, the UMATHT subroutine is used to solve the thermal constitutive behaviour of materials during transient heat transfer processes. Hence, it is necessary to adapt the framework such that temperature gradients and transient effects may be studied synergistically with H transport. This requires that a separate finite element model is developed to first solve the thermal transport problem. It is important to note that the total step time of the thermal model should match that of the H transport model. Also, model dimensions should match and ideally have the same mesh (coupled temperature-displacement elements can be used in both instances). However, matching nodal coordinates and corresponding solution output times are not required, as the software will automatically interpolate in space and time when reading nodal temperature data. Of course, it is also important that nodal temperatures are written to the output database file. If using coupled temperature displacement elements in \textit{Abaqus}, nodal temperatures are stored automatically. Temporal and spatial temperature data may then be imported to the H transport model as a predefined field variable. Moreover, for modelling thermal expansion, \textit{Abaqus'} built-in model cannot be used since it relies on the temperature increment as an input. In this approach, the temperature increment is first computed in UMATHT and then stored as a state variable that is accessed by UEXPAN to compute and impose the thermal eigenstrains. It is important to note that this decoupled approach requires that thermal transport is unaffected by H transport. This was justified in Section \ref{framework}. For models linking H transport with e.g. cracking, a fully coupled method would be required. In this paper, we only consider H redistribution in materials subjected to thermal and elastic loading.

\section{Case studies and results} \label{results}

\subsection{Case studies in iron and nickel: application to a heat exchanger geometry} \label{heatexchanger}

\subsubsection{Parallel flow heat exchanger geometry and boundary conditions}

\begin{figure}[ht]
\centering
\includegraphics[width=0.9\linewidth]{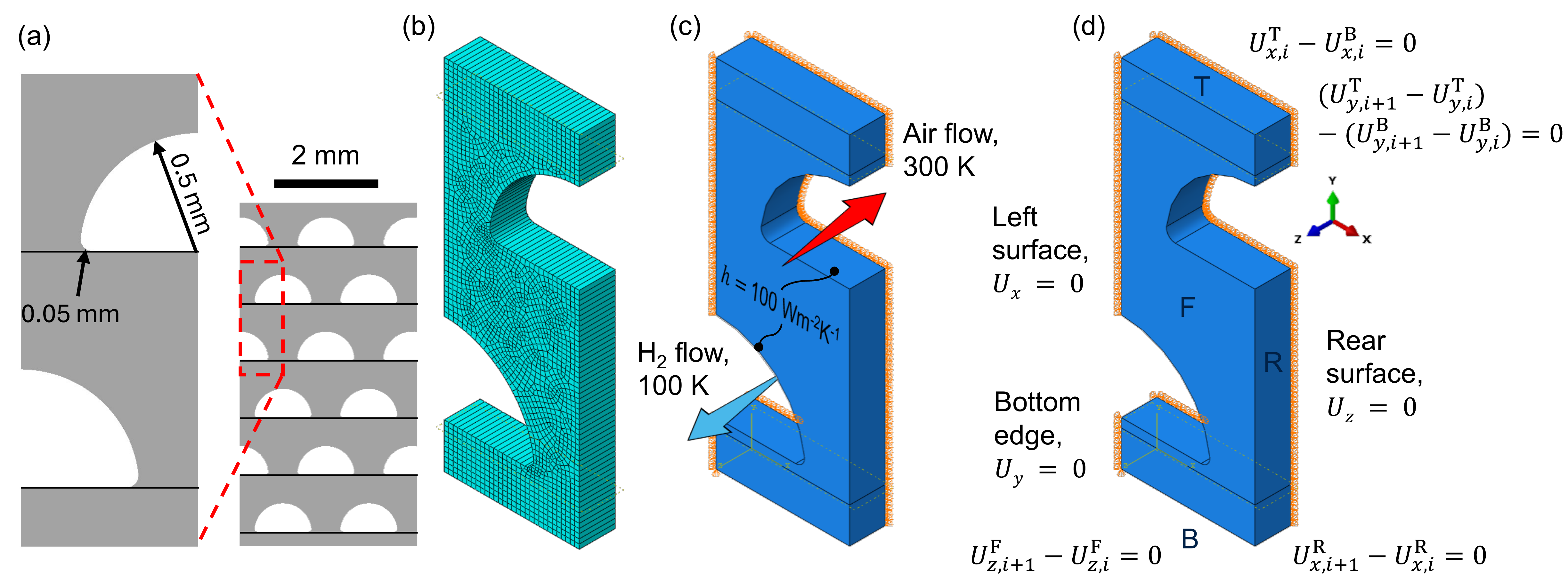}
\caption{\label{HE_BCs} Overview of heat exchanger finite element model. (a) The chosen counterflow heat exchanger geometry and unit cell. Horizontal black lines correspond to boundaries between plates (diffusion bonding lines). (b) 3D unit cell geometry and mesh. A single element is used in the through thickness direction as generalised plane strain conditions are enforced. (c) Thermal model boundary conditions. Convective heat transfer is applied to internal wall surfaces. Periodic temperature boundary conditions are enforced between top and bottom surfaces. There is no heat transfer through any other surface due to symmetry. (d) Displacement boundary conditions in H transport model. Left and rear surfaces fixed in normal directions. Generalised plane strain enforced by symmetry on right and front surfaces. Normal displacements mapped between top and bottom surfaces. Superscripts B, F, R, and T represent bottom, front, right, and top surfaces, respectively.   }
\end{figure}

In this section, we present the model used to capture coupled thermomigration and stress-driven H transport in iron and nickel heat exchangers. As shown in Figure \ref{HE_BCs} (a), a diffusion bonded (plate) counterflow heat exchanger geometry \cite{NESTELL2015} is considered, and is represented by a repeating unit cell. The geometry is modelled in 3D to enforce generalised plane strain using equation constraints between front surface nodes, with out-of plane displacements set to zero at the rear surface nodes, to keep the front and rear surfaces parallel during deformation. Generalised plane strain sections in \textit{Abaqus} do not enforce that out of plane deformation remains parallel to the original 2D plane. Figure \ref{HE_BCs} (b) shows the finite element mesh. There is no discretisation in the out-of-plane direction (z-axis), since displacements and temperatures are treated as constant in this direction. Figure \ref{HE_BCs} (c) shows the thermal model boundary conditions. Convective heat transfer is applied to both channel sections to represent heat transfer from air at 300 K to the heat exchanger, and from the heat exchanger to gaseous H at 100 K. Newton's law of cooling gives the surface heat flux along the normal direction, $\boldsymbol{n}$,

\begin{equation}\label{Jq2}
    \boldsymbol{J}_{\mathrm{q}}\cdot\boldsymbol{n}=h(T_{\mathrm{s}}-T_{\mathrm{env}}),
\end{equation}

where $h$ is the convective heat transfer coefficient, $T_{\mathrm{s}}$ the surface temperature, and $T_{\mathrm{env}}$ the environment (air or H) temperature. For these didactic case studies, temperatures are chosen to be representative of flow at a point along the heat exchanger. For each flow, a convective heat transfer coefficient of $h=$ 100 Wm$^{-2}$K$^{-1}$ is approximated \cite{QIU2023}. Figure \ref{HE_BCs} (d) shows the mechanical boundary conditions, which are implemented in the H transport model.

Symmetry between left and right surfaces (surfaces with normal along the x-axis) is enforced by setting the out-of-plane displacements to zero and to be constant (generalised plane strain), respectively. As shown, periodic displacement boundary conditions are enforced between top and bottom surfaces using equation constraints. Periodic temperature boundary conditions are also enforced between top and bottom sources. All other surfaces are treated as adiabatic due to symmetry. For each iron and nickel case study, timescales for H transport modelling are set according to their respective diffusivities to ensure a steady state H concentration profile is reached. In each case, thermal transport reaches a steady state before H transport does. The H diffusivity is assumed to exhibit an Arrhenius dependence on temperature and is given by

\begin{equation}\label{DL}
    D_{\mathrm{L}} = D_0 \exp\left(-\frac{Q_{\mathrm{D}}}{RT}\right),
\end{equation}

where $D_0$ denotes the diffusion pre-exponential factor and $Q_{\mathrm{D}}$ is the activation energy for H diffusion.
Here we do not consider the uptake of H through surfaces; we investigate the redistribution of an existing (spatially uniform) H concentration under thermigration and stress-migration driving forces. In iron and nickel, prototypical solubility concentrations (solubility lattice occupancies) for body-centred cubic and face-centred cubic materials, respectively ($\theta_{\mathrm{L}}=$ 1 ppm and 1000 ppm \cite{SUGIMOTO2014,LEE2015}; here, ppm refers to the number of H atoms per available lattice site), are assigned at $t=$ 0 s. Isotropic elasticity is assumed throughout, using property data from the literature. The material properties used in this section are given in Table \ref{Fe_Ni_props}. Thermal conductivity is generally a temperature-dependent property. In this study, however, it is assumed to be constant. This simplification is justified because the simulations for iron and nickel are conducted over a narrow temperature range, within which their thermal conductivity can be reasonably approximated as constant.

\begin{table}[htb]
\centering
\caption{Elastic, thermal transport, and H transport properties in iron and nickel. Some temperature dependent properties are approximated as constant across the temperature range considered in this study. }
\label{Fe_Ni_props}
\begin{tabular}{l l l l l}
\hline
Property & Value (iron) & Value (nickel) & Units & Source \\
\hline
$E$ & $215-0.14T\exp\left(-\frac{813}{T}\right)$ & $233-0.08T\exp\left(-\frac{173}{T}\right)$ & GPa & \cite{LI2019} \\
$\nu$ & $0.26$ & $0.32$ & -- & \cite{LEDBETTER1973} \\
$\alpha_{\mathrm{th}}$ & $14.5\times10^{-6}$ & $11.0\times10^{-6}$ & K$^{-1}$ & \cite{STRAUMANIS1969,KOLLIE1977} \\
$k$ & $87.1$ & $82.5$ & Wm$^{-1}$K$^{-1}$ & \cite{BACKLUND1961,THOMPSON1958} \\
$c_{\mathrm{p}}$ & $384.93$ & $383.06$ & Jkg$^{-1}$K$^{-1}$ & \cite{AUSTIN1932,MESCHTER1981} \\
${V}_{\mathrm{L}}$ & $2.10\times10^{-6}$ & $8.43\times10^{-7}$ & m$^3$mol$^{-1}$ & \cite{WAGENBLAST1971,THOMAS1983} \\
$Q^*_{\sigma_{\mathrm{H}}=0}$ & $-0.2T^2+367T-194420$ & $0.01T^2+20T-30192$ & Jmol$^{-1}$ & This work \\
$D_0$ & $4.20\times10^{-8}$ & $4.47\times10^{-7}$ & m$^2$s$^{-1}$ & \cite{NAGANO1982,HILL1955} \\
$Q_{\mathrm{D}}$ & $3850$ & $35982$ & Jmol$^{-1}$ & \cite{NAGANO1982,HILL1955} \\
\hline
\end{tabular}
\end{table}

\subsubsection{Competing effects of stress-driven hydrogen transport and thermomigration}

\begin{figure}[ht]
\includegraphics[width=1\linewidth]{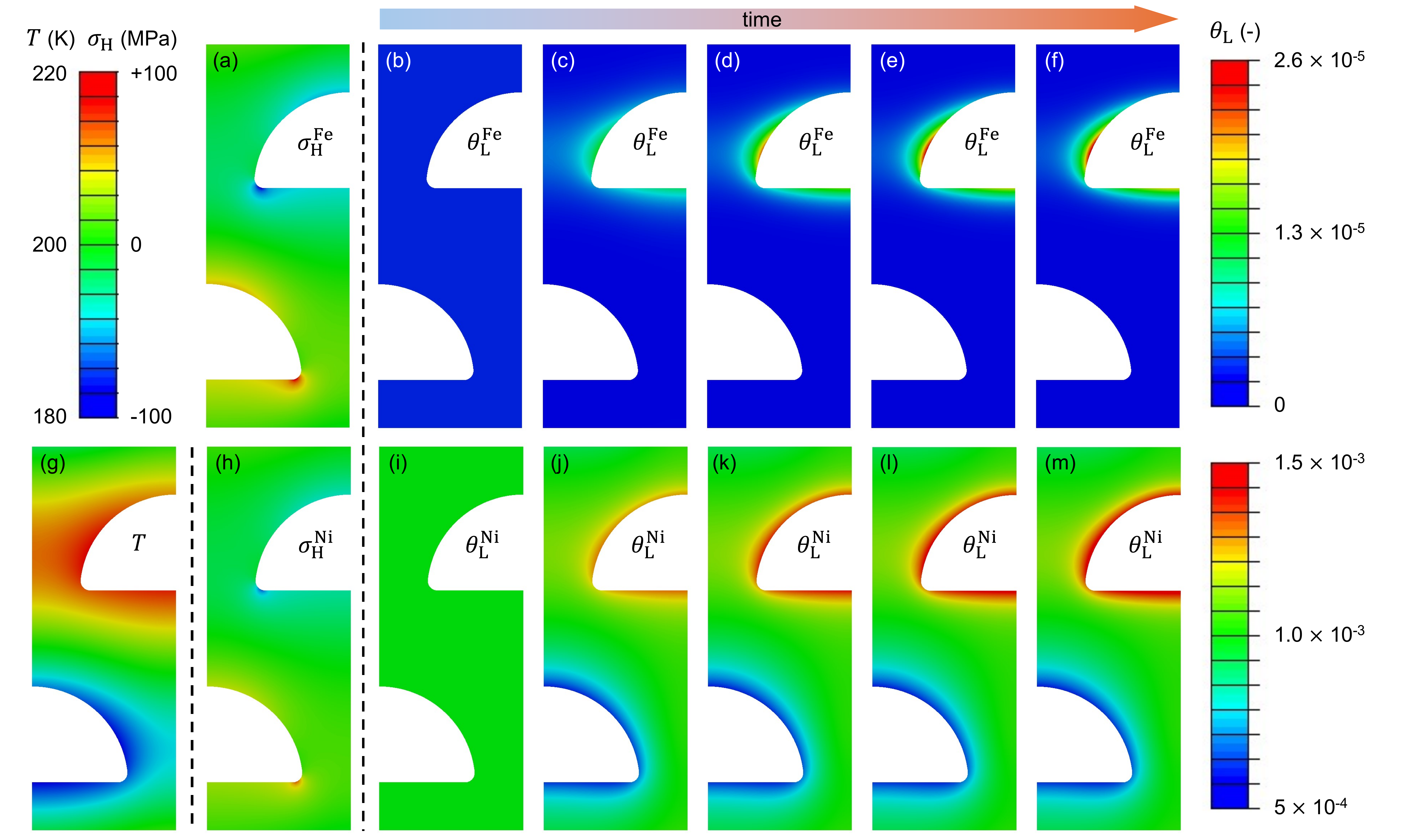}
\caption{\label{Fe_Ni_contours}  Distributions of (a) stress at steady state and (b) - (f) lattice occupancy in iron at $t=0$, $25$, $50$, $75$, and $100$ s, respectively. The steady state temperature profiles in iron and nickel are near equivalent and are shown in (g). Lattice occupancy distributions in nickel are shown in (h) - (i) at $t=0$, $7.5\times10^5$, $1.5\times10^6$, $2.25\times10^6$, and $3\times10^6$ s, respectively.}
\end{figure}

Figure \ref{Fe_Ni_contours} shows the distributions of hydrostatic stress, temperature, and lattice occupancy in iron and nickel heat exchanger cells. Steady state hydrostatic stress distributions are shown in sub-figures (a) and (h). The lower corners of the heat exchanger channels, corresponding to interfaces between diffusion-bonded layers, are shown to give rise to stress concentrations in each case. The temperature field driving these incompatibility stresses is shown in sub-figure (g). A single temperature field is shown at $t=100$ s as the thermal transport properties for iron and nickel are very similar over the temperature range considered here (see Table \ref{Fe_Ni_props}). Tensile stresses are shown to concentrate around the corner of the cold channel (where there is H flow), while compressive stresses develop around the corner of the hot channel (where there is air flow). In nickel, peak stresses are shown to be lower than they are in iron, due to its lower thermal expansion coefficient. 

Lattice occupancy distributions are shown through time from $t=0$ to steady state in iron and nickel in sub-figures (b) - (f) and (i) - (m), respectively. Owing to large differences in diffusivity, the timescales for H transport in iron and nickel differ dramatically: in iron, a steady state is achieved within minutes, while in nickel, this happens on the order of one month. In the iron example, H is shown to redistribute significantly from an initial uniform lattice occupancy of 1 ppm to upwards of 26 ppm around the hot channel at steady state. This substantial redistribution results in an effective clearing of H from around the cold channel, where the lattice occupancy reduces to as low as 10$^{-6}$ ppm. In materials with a negative heat of transport (as is the case for iron and nickel in this temperature range), thermomigration drives H from low- to high-temperature regions. Similarly, the stress-migration driving force promotes H migration from low to high hydrostatic stress regions. Hence, thermomigration and stress-driven H transport are competing in this analysis. In the iron example, thermomigration is evidently the dominant driving force, as there is no evidence for H concentrating around the high tensile stress region.

A similar conclusion can be drawn from the nickel analysis, as the H population is shown to concentrate around the hot channel. It differs from the iron case, however, as redistribution is somewhat more symmetric; H concentration increases by $\approx$ 50\% around the hot channel and decreases by $\approx$ 50\% around the cold channel. Moreover, levels of redistribution are significantly lower in nickel with respect to the initial H population (it was shown analytically in \cite{LONG2025} that the level of H redistribution, normalised by the initial concentration, is independent of the initial concentration). In zones with high tensile and compressive hydrostatic stress, there is no evidence here of associated H localisation or displacement. Hence, thermomigration is also clearly dominant in the nickel heat exchanger, though the overall level of H redistribution is much lower than in iron, mainly owing to differences in heat of transport magnitude. Despite the heat of transport magnitude being lower in nickel, thermomigration remains the dominant driving force. This is partly explained by the relatively low thermal expansion coefficient in nickel, leading to lower overall hydrostatic stress gradients. Furthermore, the partial molar volume of H in nickel is lower than it is in iron by more than a factor of two (which, incidentally, is what governs the vastly different H solubility limits in the two materials \cite{ORIANI1994}). 

\subsubsection{An efficient graphical method for engineering analysis}

As the competition between thermomigration and stress-driven H transport arises from a complex interplay between temperature gradients, hydrostatic stress gradients, and material properties, the resulting transport behaviour is often unintuitive. Although advanced H transport simulations, such as those presented in the previous subsection, can accurately capture this behaviour, practical engineering design often benefits from more streamlined approaches. Here we introduce an alternative modelling approach that enables rapid identification of the dominant driving force(s) and the resulting direction of H transport. Starting from the constitutive law for H transport given in Equation~(\ref{JH1}), we consider the limiting case in which the thermomigration and stress-migration driving forces contribute equally to the H flux. Assuming the absence of a H concentration gradient, this condition corresponds to zero net H flux and may be expressed as $-V_{\mathrm{L}} \nabla \sigma_{\mathrm{H}} + \frac{Q^*}{T} \nabla T = 0$. Rearranging this expression, and extending it to situations in which thermomigration and stress-migration act synergistically rather than in opposition, yields the following approximate relation for the ratio of the hydrostatic stress gradient to the temperature gradient,

\begin{equation}\label{ratio}
\left| \frac{\nabla \sigma_{\mathrm{H}}}{\nabla T} \right|_{\boldsymbol{\bar{J}}_{\mathrm{H}} = 0}
\approx
\left| \frac{Q^*_{\sigma_{\mathrm{H}} = 0}}{V_{\mathrm{L}} T} \right|.
\end{equation}

Equation (\ref{ratio}) represents an approximate solution, obtained by adopting the convenient approximation $Q^* \approx Q^*_{\sigma_{\mathrm{H}} = 0}$, which is justified by the observation that the contribution of hydrostatic stress to $Q^*$ is often negligible \cite{LONG2025THERMO}. This relation provides a convenient criterion for assessing the dominant driving force for H transport. In particular, comparison of the magnitudes of the hydrostatic stress and temperature gradients yields

\begin{equation}
\left| \dfrac{\nabla \sigma_{\mathrm{H}}}{\nabla T} \right| 
> \left| \dfrac{Q^*}{V_{\mathrm{L}} T} \right| 
\quad \Rightarrow \quad \text{stress-migration dominated},
\end{equation}

\begin{equation}
\left| \dfrac{\nabla \sigma_{\mathrm{H}}}{\nabla T} \right| 
< \left| \dfrac{Q^*}{V_{\mathrm{L}} T} \right| 
\quad \Rightarrow \quad \text{thermomigration dominated}.
\end{equation}

\begin{figure}[ht]
\includegraphics[width=1\linewidth]{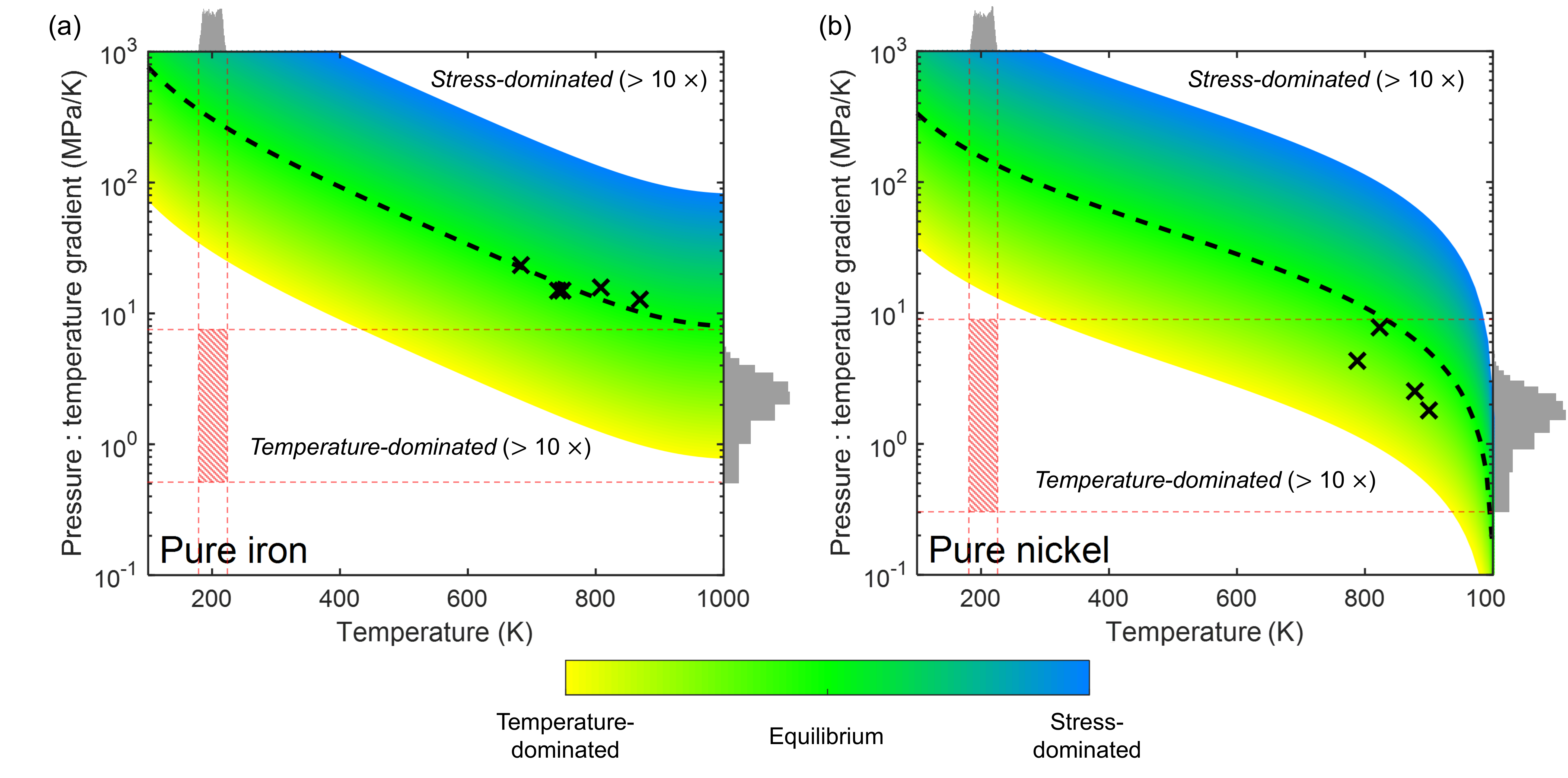}
\caption{\label{Fe_Ni_graph} Graphical analysis of contributions from stress and temperature gradients to H transport in (a) iron and (b) nickel. The black dashed line represents the hydrostatic stress to temperature gradient ratio at which the contributions to H transport are equal in magnitude (based on Equation (\ref{Q*1}) for $Q^*$). Individual markers represent the same metric, but use real $Q^*$ measurements \cite{GONZALEZ1965} directly. The shaded region represents the domain in which either stress or temperature dominates by up to one order of magnitude. Histograms along horizontal and vertical axes represent the predicted steady state distributions of temperature and hydrostatic stress to temperature gradient ratio from the analyses in Section \ref{Fe_Ni_contours}. The intersections of these distributions are indicative of the H transport regime (e.g. temperature-dominated, stress-dominated, or multimodal).  }
\end{figure}

By plotting Equation (\ref{ratio}) as a function of temperature, a threshold condition separating thermomigration-dominated and stress-migration-dominated H transport can be identified. This condition is represented in Figure~\ref{Fe_Ni_graph} by a black dashed line. On either side of the dashed line, a transition zone is identified in which one driving force dominates over the other by up to one order of magnitude. Outside this zone, H transport is clearly dominated by either thermomigration or stress-driven migration, respectively. In sub-figures (a) and (b), the distributions of temperature and of the hydrostatic stress (pressure) gradient to temperature gradient ratio are shown as frequency distributions along their respective axes. Unlike the H transport results presented in Figure \ref{Fe_Ni_contours}, gradients of hydrostatic stress and temperature are easily attained from simple elastic and thermal transport analyses. Using the intersections of the frequency distributions, a process window for each case study can be readily established. As shown in Figure~\ref{Fe_Ni_contours}, the process windows for the iron and nickel case studies lie well within the temperature-dominated domain, as expected. Consequently, given knowledge of the stress gradient direction relative to the temperature gradient (typically opposite in thermal incompatibility problems) and of the sign of the heat of transport (which governs the direction of thermomigration), engineers may use this graph-based approach to reliably assess the net direction of H transport in complex coupled problems such as this. This method is particularly valuable in cases where thermomigration and stress-migration compete, but it may also inform design decisions in scenarios where the two mechanisms act synergistically and one clearly dominates.

For highly constrained thermal incompatibility problems such as this, further coarse approximations may be used to roughly identify a characteristic hydrostatic stress gradient to temperature gradient ratio. Incompatibility stresses arise according to $\sigma_{ii}=E\varepsilon_{ii}^{\mathrm{el}}=-E\varepsilon_{ii}^{\mathrm{th}}=-E\alpha_{\mathrm{th}}\Delta T$, where $\varepsilon_{ii}^{\mathrm{el}}$ are the principal elastic strains, $\varepsilon_{ii}^{\mathrm{th}}$ the principal thermal strains, and $\alpha_{\mathrm{th}}$ is the coefficient of thermal expansion. Assuming a three-dimensional body that is fully constrained in all three principal directions, the hydrostatic stress may be written as $\sigma_{\mathrm{H}} = \frac{1}{3}\sum_{i=1}^{3}\sigma_{ii} = - E \alpha_{\mathrm{th}}\Delta T$. Taking the spatial gradient of the hydrostatic stress yields $\nabla \sigma_{\mathrm{H}} = - E \alpha_{\mathrm{th}} \nabla T$, from which a characteristic ratio of the hydrostatic stress gradient to the temperature gradient may be obtained as $\left| \frac{\nabla\sigma_{\mathrm{H}}}{\nabla T} \right|\approx
E \alpha_{\mathrm{th}}$. For iron and nickel, this yields values of around 3.1 and 2.5 MPa K$^{-1}$, respectively, which are consistent with the distributions shown in Figure \ref{Fe_Ni_graph}. This estimate suggests that, in highly constrained geometries without pronounced stress concentrators, thermomigration will generally dominate in Fe and Ni. Only in the presence of strong stress raisers (e.g., sharp channel corners, crack tips) or externally applied mechanical loads would stress gradients significantly exceed the bound set by $E\alpha_{\mathrm{th}}$.

An important implication is that the product $E\alpha_{\mathrm{th}}$ provides a first-order, material-specific metric for assessing the relative importance of thermally induced stress gradients. In this simplified fully constrained limit, whether temperature- or incompatibility stress gradients dominate H redistribution appears largely governed by material properties. However, this interpretation must be treated with caution: the actual stress state depends strongly on geometric constraint, boundary conditions, and external loads. Relaxation mechanisms, partial constraint, and multiaxial stress states can substantially alter the effective hydrostatic stress relative to the fully constrained estimate. Therefore, while $E\alpha_{\mathrm{th}}$ offers a useful screening parameter for rapid design assessment, detailed numerical modelling remains necessary when constraint conditions or mechanical loading are non-negligible.

\subsection{Application of an efficient graphical method to zirconium alloy fuel cladding} \label{Zircaloy}

\subsubsection{Nuclear fuel cladding model}

Another prominent industrial example of coupled H thermomigration and stress-migration occurs in nuclear power applications. In pressurised water reactors, zirconium alloy cladding materials are used to contain nuclear fuel pellets while facilitating heat transfer from the fuel to the coolant. Owing to the small cladding wall thicknesses, typically on the order of 0.5 mm, very large temperature gradients are established across the cladding. These gradients give rise to thermal incompatibility stresses along the length of the cladding. At the external cladding surface, oxidation in high-temperature water produces H electrochemically, which is then absorbed into the material. A critical concern for reactor operation is that H concentrations may become sufficiently high to precipitate brittle zirconium hydrides, leading to severe degradation of the cladding’s structural integrity. Hence, accurate predictive structural modelling of cladding performance requires that the active H transport mechanisms be fully and consistently accounted for.

\begin{figure}[ht]
\centering
\includegraphics[width=0.6\linewidth]{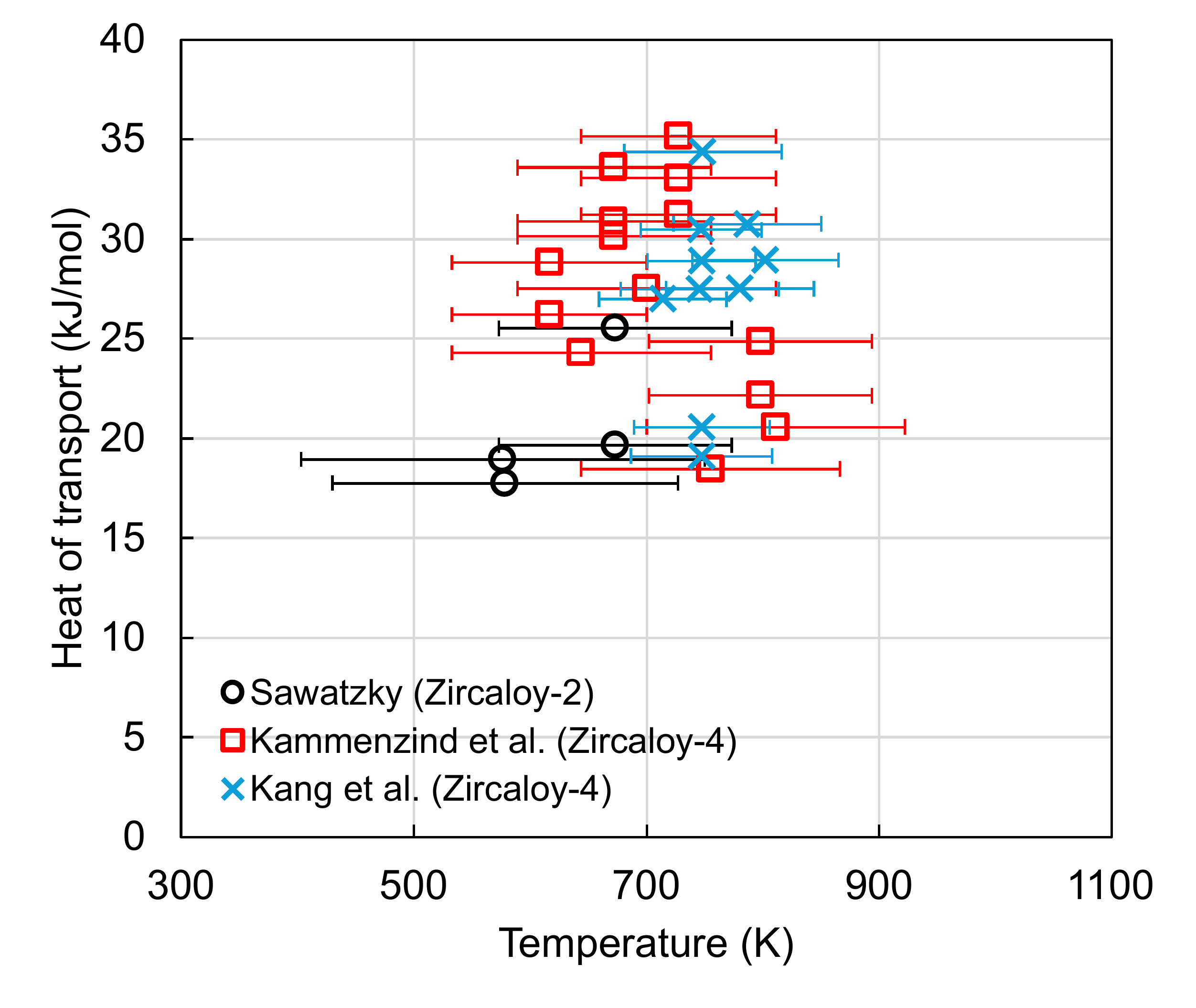}
\caption{\label{Zr_Qstar}  Published experimental measurements \cite{SAWATZKY1960,KAMMENZIND1996,KANG2023} of $Q^*$ in zirconium fuel cladding alloys (Zircaloy-2 and Zircaloy-4) at different temperatures. The temperature range (as per the temperature gradient) for each measurement is shown using horizontal error bars.  }
\end{figure}

A recurring challenge in modelling H thermomigration is the scarcity of reliable experimental data for the heat of transport. Many studies focus on zirconium alloys, where experimental approaches are comparatively simpler than in iron and nickel, for example. Owing to hydride precipitation, heat of transport measurements in zirconium alloys do not employ permeation configurations such as those used by Gonzalez and Oriani \cite{GONZALEZ1965}. Instead, specimens are pre-charged with H and subjected to a thermal gradient, allowing H redistribution by thermomigration. Subsequent rapid cooling promotes zirconium hydride precipitation, effectively `freezing' the H distribution for post-mortem sectioning and measurement. While this approach is experimentally straightforward, it is also associated with significant scatter and uncertainties related to hydride nucleation, growth kinetics, and the assumption that the measured hydride distribution faithfully represents the high-temperature H profile. Figure \ref{Zr_Qstar} compiles reported heat of transport values for Zircaloy-2 and Zircaloy-4 (zirconium cladding alloys), showing substantial scatter between studies \cite{SAWATZKY1960,KAMMENZIND1996,KANG2023}. This variability reflects known limitations of the hydride-based approach. The data are shown with horizontal error bars that represent the large temperature ranges inherent to these experiments, arising from the use of relatively long specimens compared with permeation specimens. Together, these factors highlight that, while experimentally simple, this methodology carries significant uncertainty and the reported $Q^*$ should be interpreted carefully.

\begin{figure}[ht]
\includegraphics[width=1\linewidth]{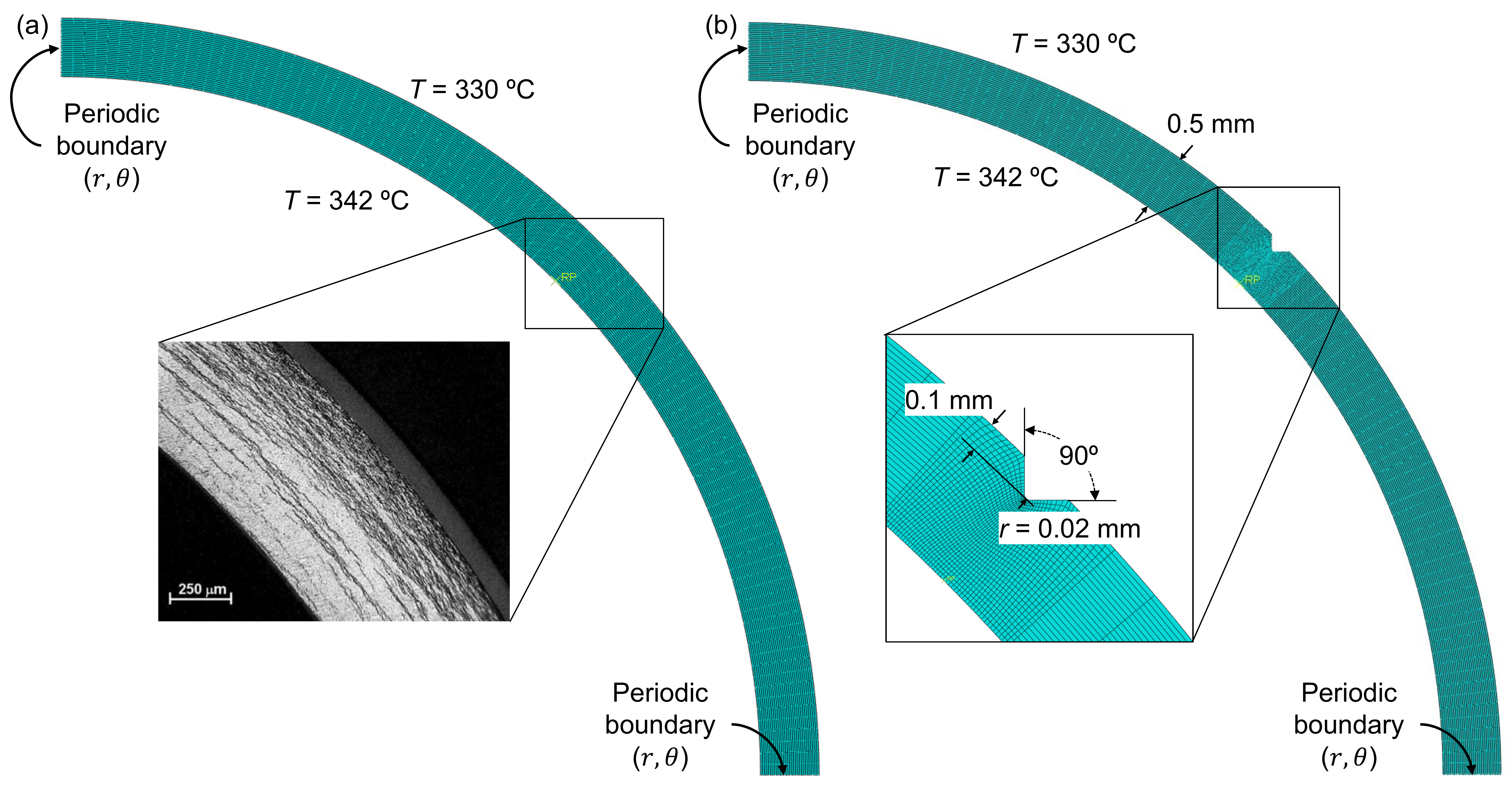}
\caption{\label{Zr_BCs} Model geometry and boundary conditions in zirconium alloy fuel cladding analyses. Four problems are presented: two in cladding with (a) uniform cross-section and two in cladding with (b) a notch at the outer surface, representative of e.g. hydride blistering damage \cite{KIM2018}. Plane strain elements are used to represent the pipe sections. In each example problem, a radial temperature gradient is established, based on representative internal and external wall temperature \cite{LAMARSH1977,MOTTA2015} boundary conditions. Radial incompatibility stress gradients arise as a consequence. In each (a) and (b), an additional problem is analysed in which an internal gas pressure of 15 MPa is applied, representative of fission gas release at high levels of burn-up \cite{EIDELPES2019}. (a) Includes a micrograph of a post-operation section of fuel cladding, showing a highly non-uniform distribution of zirconium hydrides, concentrated near the outer surface. Micrograph reproduced from U.S. Nuclear Regulatory Commission NUREG 6967 \cite{BILLONE2008}.}
\end{figure}

Though we have developed a mechanistic model for $Q^*$, many key input parameters (such as $\frac{\partial\rho}{\partial N_{\mathrm{H}}}$) are unavailable for zirconium and its alloys. This limitation largely stems from hydride precipitation, which makes it extremely difficult to isolate properties associated with mobile lattice H. In this section, we present a coupled thermal and mechanical model of a zirconium alloy cladding section subjected to a temperature gradient. Four case studies are considered, each with different stress distributions resulting from: (i) thermal incompatibility, (ii) thermal incompatibility combined with internal pressure, (iii) thermal incompatibility with a stress-raiser, and (iv) thermal incompatibility combined with internal pressure and a stress-raiser. Due to the non-availability of a $Q^*$ model for zirconium, we use the graphical method presented in the previous section to analyse the results of each case study, relying on experimental $Q^*$ measurements only. Geometries and boundary conditions for these cases are summarised in Figure \ref{Zr_BCs}.

Figure \ref{Zr_BCs} (a) shows the 2D (plane strain) quarter-model geometry of the cladding used in case studies (i) and (ii). A micrograph of a post-operation cladding section \cite{BILLONE2008} is also shown, highlighting pronounced hydride precipitation near the outer surface. This observation indicates that the dominant driving forces for H transport promote elevated H concentrations in this region. Figure \ref{Zr_BCs} (b) shows the notched geometry employed in case studies (iii) and (iv). The same mechanical loading and thermal boundary conditions as in cases (i) and (ii) are applied. The notch is introduced to amplify hydrostatic stress gradients relative to the imposed temperature gradient, thereby enabling the exploration of a broad range of H transport conditions. This geometric feature is representative of oxidation-induced damage and hydride blistering commonly observed in cladding materials during in-service operation \cite{MOON1970,KIM2018}. Isotropic linear elasticity is assumed throughout. Although zirconium alloys exhibit crystallographic anisotropy at the grain scale, the cladding material considered here is treated as mechanically isotropic at the continuum level due to its fine-grained, polycrystalline microstructure. The crystallographic texture of zirconium claddings is not unique and varies with alloy composition and thermomechanical processing route \cite{RODRIGUEZ1992}. Moreover, the primary focus of this study is on relative H redistribution driven by coupled thermal and hydrostatic stress gradients, rather than on local anisotropic elastic responses. The elastic properties used in the simulations are summarised in Table \ref{Zr_props}.

\begin{table}[htb]
\centering
\caption{Elastic, thermal transport, and H transport properties in zirconium. Some temperature dependent properties are approximated as constant across the temperature range considered in this study.}
\label{Zr_props}
\begin{tabular}{l l l l}
\hline
Property           & Value                            & Units              & Source               \\
\hline
$E$                & $-0.031T+106$                    & GPa                & \cite{WHITMARSH1962} \\
$\nu$              & $0.44$                           & --                 & \cite{WHITMARSH1962} \\
$\alpha_{\mathrm{th}}$           & $3\times10^{-9}T+5\times10^{-6}$ & K$^{-1}$           & \cite{WHITMARSH1962} \\
$k$                & $14.1$                           & Wm$^{-1}$K$^{-1}$  & \cite{WHITMARSH1962} \\
$c_{\mathrm{p}}$   & $0.113T+262$                     & Jkg$^{-1}$K$^{-1}$ & \cite{WHITMARSH1962} \\
${V}_{\mathrm{L}}$ & $1.66\times10^{-6}$              & m$^3$mol$^{-1}$    & \cite{EADIE1992}     \\
\hline
\end{tabular}
\end{table}

\subsubsection{Synergistic coupling of stress-driven hydrogen transport and thermomigration}

\begin{figure}[ht]
\includegraphics[width=1\linewidth]{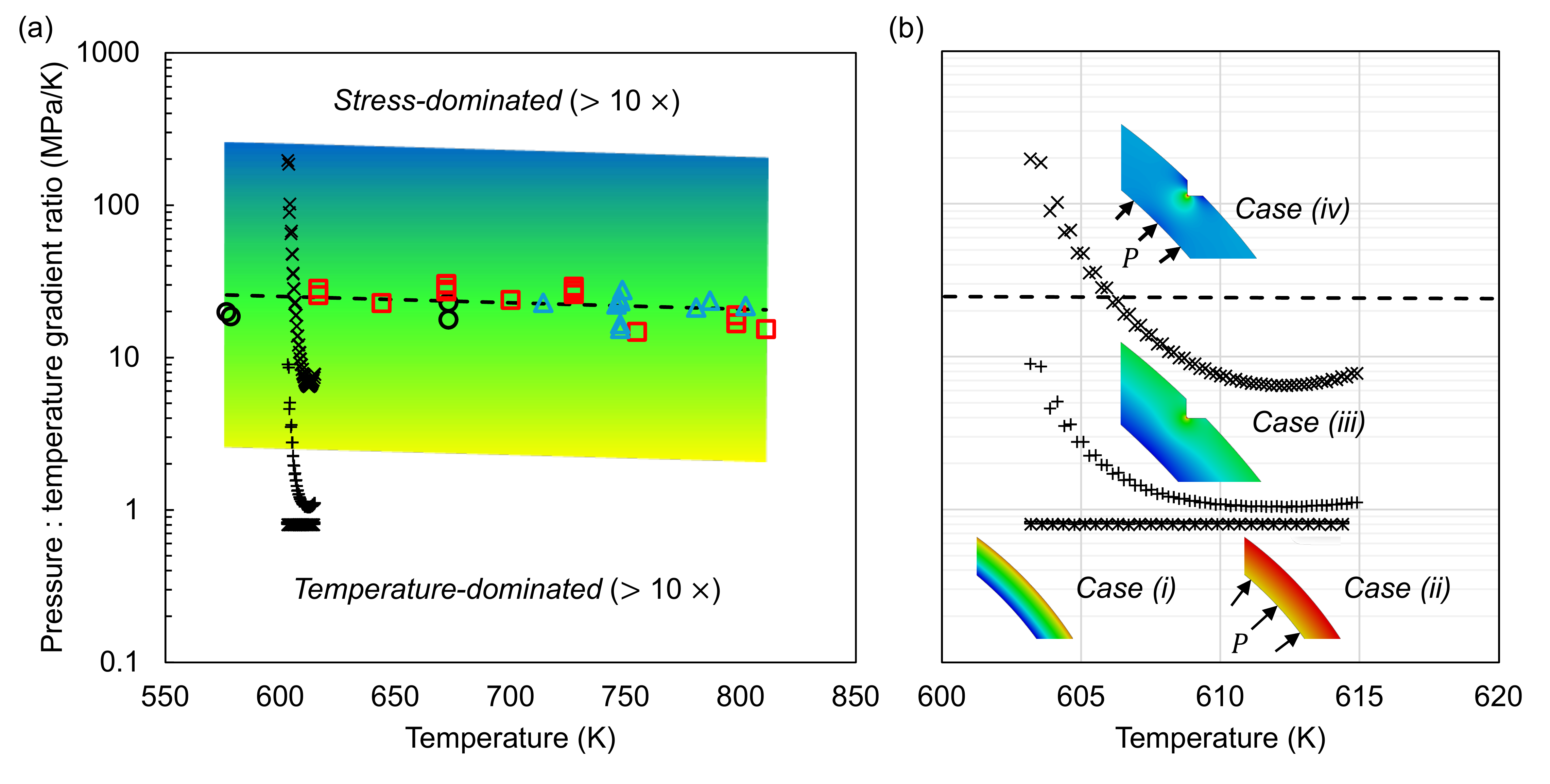}
\caption{\label{Zr_graph} (a) Graphical analysis of contributions from stress and temperature gradients to H transport in zirconium alloy cladding. Heat of transport data shown in Figure \ref{Zr_Qstar} are used to approximate a region in which stress and thermomigration driving forces make equal contributions. The shaded region represents variation by up to one order of magnitude in either direction, as before. Predicted pressure to temperature gradient ratio distributions (radially from outer to inner surfaces) from cases (i) to (iv) are overlaid. In notched examples, distributions are plotted radially from the notch root. (b) Shows these distributions across a narrower temperature range. Hydrostatic stress contour plots are shown for illustration.  }
\end{figure}

As a result of thermal expansion, hydrostatic stress gradients arising from thermal incompatibility develop with a sign opposite to that of the temperature gradient. These stress gradients therefore promote H transport from regions of high temperature to low temperature, corresponding to the outward radial direction. Moreover, unlike iron and nickel, zirconium alloys reportedly exhibit a positive $Q^*$, such that H thermomigration also drives H from high- to low-temperature regions. As a result, both mechanisms act synergistically to produce a net outward radial H flux, consistent with the micrograph shown in Figure \ref{Zr_BCs} (a). This raises the question of which effect dominates under what conditions? In Figure \ref{Zr_graph}, we use the graph-based approach to study cases (i) to (iv). 

Figure \ref{Zr_graph} (a) presents the distributions of the ratio between the hydrostatic stress gradient and the temperature gradient for cases (i) to (iv). Figure \ref{Zr_graph} (b) provides a magnified view of these distributions, corresponding to a one-dimensional profile along the inward radial direction from the external (low-temperature) to the internal (high-temperature) surface. For the notched cases, the profiles are plotted from the notch root. In sub-figure (a), a best-fit line is used to approximate the temperature dependence of the term defined in Equation (\ref{ratio}), corresponding to the condition under which the two H transport driving forces contribute equally. The shaded region surrounding this line indicates the domain in which one driving force dominates over the other by up to one order of magnitude. Both cases (i) and (ii) lie clearly within the temperature-dominated regime. Although the applied internal pressure in case (ii) marginally increases the hydrostatic stress gradient, its overall influence on the ratio is negligible. Moreover, the ratio of gradients is shown to be effectively independent of temperature. The earlier approximation derived for thermal incompatibility problems gives $\left|\frac{\nabla \sigma_{\mathrm{H}}}{\nabla T}\right| \approx E \alpha_{\mathrm{th}} = 0.6$ MPa/K when applied here. This is in close agreement (within 10\%) with the finite element results for case (i). For simple problems of this type, this coarse approximation provides a fast and reliable estimate of the relative magnitudes of the thermomigration and stress-driven transport contributions. In contrast, for cases (iii) and (iv), the notched geometry introduces a pronounced stress concentration that shifts the transport behaviour towards stress dominance. This is particularly clear for case (iv), with internal pressure, showing stress-dominance of up to one order of magnitude in the vicinity of the notch. 

\section{Discussion}

The results presented in this paper highlight that the competition between thermomigration and stress-driven H transport is highly system-dependent and governed by the relative magnitudes of temperature gradients, hydrostatic stress gradients, and material-specific transport parameters. While the case studies considered here focus on internal H redistribution under prescribed thermal and mechanical loading, real engineering components are further complicated by surface-mediated H uptake and evolving boundary conditions. In applications such as heat exchangers and nuclear fuel cladding, H is not only redistributed internally but is continuously absorbed from and desorbed to the surrounding environments. Accurately predicting service-relevant H concentrations therefore requires coupling internal transport mechanisms, as described here, with H absorption boundary conditions. This challenge lies beyond the scope of the present work but is critical for future model development.

In the heat exchanger case studies, thermomigration is shown to dominate H redistribution in both iron and nickel, despite the presence of significant thermal incompatibility stresses. An important and somewhat counter-intuitive implication of this finding is that dominant thermomigration drives H away from regions of high tensile stress, which are typically associated with HE. While this redistribution may locally reduce embrittlement risk, it may simultaneously promote more rapid net H uptake by sustaining a high net inward H flux from exposed surfaces. Hence, thermomigration-dominated transport may indirectly lead to more rapid and greater total H absorption, even while mitigating localised stress-driven accumulation. This trade-off underscores the importance of considering global H transport, rather than local concentrations alone, when assessing the structural integrity of components.

The zirconium alloy cladding analyses demonstrate that geometric stress concentrators can significantly alter the balance between transport mechanisms. While smooth, axisymmetric cladding sections remain clearly within the temperature-dominated regime, the introduction of notches produces steep hydrostatic stress gradients that dominate over thermomigration locally. This finding reinforces the expectation that true stress intensification (crack tips) will always dominate over thermomigration in their immediate vicinity. As a result, the framework presented here is particularly well suited to analysing conditions relevant to delayed hydride cracking (DHC) \cite{SINGH2002}, where stress-driven H accumulation ahead of crack tips governs crack initiation and growth. A key open question emerging from this work is the existence of a critical defect size beyond which stress-driven transport becomes dominant, triggering DHC. The proposed graphical method offers a practical route for identifying such thresholds without resorting to fully coupled, high-fidelity simulations.

The case studies presented here adopt isotropic transport and elasticity assumptions, which are appropriate for this didactic paper. For more complete studies however, particularly for textured zirconium alloys, more careful attention should be paid to these details. Strong crystallographic texture may introduce directional dependence in both stress-driven H transport and overall lattice diffusivity. In such cases, hydrostatic stress alone may be insufficient to characterise the driving force for H migration, and deviatoric or orientation-dependent effects may become relevant \cite{LIU2025}. Extending the present framework to account for anisotropic transport tensors therefore represents an important avenue for future work. 

Overall, this study demonstrates that while thermomigration often governs H redistribution in thermally loaded components, stress-driven transport becomes decisive in the presence of steep stress gradients. The graphical approach introduced here provides a computationally efficient means of identifying these regimes, offering valuable guidance for both component design and failure assessment in H-exposed systems. 

\section{Conclusions}

We have developed a thermodynamically consistent modelling framework to quantify and compare the competing contributions of thermomigration and stress-driven transport to H redistribution under non-isothermal conditions. The framework was demonstrated through heat exchanger and nuclear fuel cladding case studies, leading to the following conclusions:
\begin{itemize}
  \item Thermomigration must be explicitly and consistently accounted for in non-isothermal problems, as it is shown to dominate H redistribution in thermal transport systems, even in the presence of substantial hydrostatic stress gradients.
  \item In materials exhibiting a negative $Q^*$, thermomigration drives H from colder to hotter regions, which may promote enhanced H uptake from cold H-containing environments, such as in heat exchanger applications.
  \item Stress-driven H transport is shown to dominate locally in the vicinity of sharp stress concentrators, such as notches, highlighting its critical role in phenomena including DHC in zirconium alloy fuel cladding.
  \item A computationally efficient graphical method has been introduced that enables rapid identification of the dominant H transport driving force, offering a practical alternative to fully coupled H transport simulations for engineering analysis and design.
\end{itemize}

\section*{Acknowledgements}

The authors would like to acknowledge Rolls-Royce plc for their financial and technical support in this project (grant number RR/UTC/89/9 BPC 189). The work was undertaken in support of the HYEST Programme (UKRI reference 10039813), funded by the Aerospace Technology Institute and UK Research and Innovation. We would particularly like to thank Chris Argyrakis, Louise Gale, Al Lambourne, Duncan Maclachlan, Rob Marshall, and Carrie Miszkowska, for their input. This work was in part supported by EPSRC program grant EP/Z534456/1.

\bibliographystyle{unsrt}
\bibliography{References}

\begin{appendices}
\appendix
\counterwithin{table}{section}
\section{Mechanistic heat of transport model properties}

\begin{table}[htb]
\caption{\label{Mechanistic_props}
Property data used in modelling the $Q^*$ \cite{LONG2025THERMO}. Data are primarily obtained from experimental measurements, with references provided where applicable. Degeneracies are estimated based on the relative intensities of peaks in the vibrational spectra.
}
\centering
\begin{tabular}{l l l l l}
\hline
Property & Value (iron) & Value (nickel) & Units & Source \\
\hline
$\bar{\omega}_i$ & $880$, $1060$ & $800$, $940$ & cm$^{-1}$ & \cite{baro1981,HOCHARD1995} \\
$g_i$ & $3$, $2$ & $3$, $2$ & (-) & \cite{baro1981,HOCHARD1995} \\
$M$ & $2$ & $2$ & (-) & \cite{baro1981,HOCHARD1995} \\
$Z_{\mathrm{es}}$ & $0.6$ & $0.25$ & $e$ & \cite{ITSUMI1996,WENG2012} \\
$\alpha_{\mathrm{S}}$ & $3.58\times10^{-5}T^2-0.07T+32.2$ & $-0.03T+1.74$ & $\mu$VK$^{-1}$ & \cite{SECCO2017,HAUPT2020} \\
$\alpha_{\mathrm{T}}$ & $7.14\times10^{-5}T^2-0.07T$ & $-0.03T$ & $\mu$VK$^{-1}$ & $T\frac{\partial\alpha_{\mathrm{S}}}{\partial T}$ \\
$\rho_{\mathrm{th}}$ & $4.00\times10^{-10}T$ & $2.26\times10^{-10}T+1.20\times10^{-7}$ & $\Omega$m & \cite{WAGENKNECHT2015} \\
$\frac{d\rho}{dN_{\mathrm{H}}}$ & $2.45\times10^{-35}$ & $7.31\times10^{-36}$ & $\Omega$m$^4$ & \cite{SINGH2025,PAPASTAIKOUDIS1983} \\
$N_{\mathrm{L}}$ & $8.49\times10^{28}$ & $9.15\times10^{28}$ & m$^{-3}$ & \cite{SMITHELLS} \\
\hline
\end{tabular}
\end{table}
\end{appendices}
\end{document}